# Hydrogen-bond equilibria and life times in a supercooled monohydroxy alcohol

C. Gainaru,[1] S. Kastner,[2] F. Mayr,[2] P. Lunkenheimer,[2] S. Schildmann,[1] H. J. Weber,[1] W. Hiller,[3] A. Loidl,[2] and R. Böhmer[1]

[1]*Fakultät für Physik, Technische Universität Dortmund, 44221 Dortmund, Germany*
[2]*Institut für Physik, Universität Augsburg, 86135 Augsburg, Germany*
[3]*Fakultät für Chemie, Technische Universität Dortmund, 44221 Dortmund, Germany*

Dielectric loss spectra covering 13 decades in frequency were collected for 2-ethyl-1-hexanol, a monohydroxy alcohol that exhibits a prominent Debye-like relaxation, typical for several classes of hydrogen-bonded liquids. The thermal variation of the dielectric absorption amplitude agrees well with that of the hydrogen-bond equilibrium population, experimentally mapped out using near infrared (NIR) and nuclear magnetic resonance (NMR) measurements. Despite this agreement, temperature-jump NIR spectroscopy reveals that the hydrogen-bond switching rate does not define the frequency position of the prominent absorption peak. This contrasts with widespread notions and models based thereon, but is consistent with a recent approach.

Hydrogen bonds are relatively weak so that thermal energies can lead to their reversible rupture and re-formation. Nature makes perfect use of the finite lifetime and the resulting flexibility of H bonds in biological systems, be it to stabilize macromolecular structures and govern their functionality or be it to cause the fascinating anomalies of their solvent, water [1]. However, even for water, it is not clear how the H bond dynamics relates to such basic properties as the prominent microwave absorption band, exploited so often every day. More closely inspected, this band features not only a profile characteristic of a single-exponential relaxation process, first quantified by Debye [2], but in addition a faint high-frequency excitation [3]. This kind of composite shape is not unique to water but shared by other H bonded liquids, notably the monohydroxy alcohols [4,5] which can easily be supercooled, enabling their study over particularly wide ranges in temperature and frequency. Thermodynamic and viscoelastic measurements permitted to identify the weak high-frequency feature as the structural relaxation, also called α-process [6,7] that governs the viscous flow and leads to a non-exponential decay of intermolecular correlations. The α-process generally represents the slowest process in small-molecule liquids [8], with monohydroxy alcohols, and by inference also water [9], as striking exceptions. The unusual, slower-than-structural Debye-like dynamics is often assigned to the relaxation of H bond mediated supramolecular structures [10]. Thus, it seems natural to develop models based on the idea that the H bond switching time scale defines the one on which the Debye-like relaxation takes place [3,4,10,11].

Applying time-dependent NIR spectroscopy to the well-studied monohydroxy alcohol 2-ethyl-1-hexanol (2E1H) the present work demonstrates experimentally that this seemingly well-established identification of time scales is not tenable. Furthermore, by comparing NIR with NMR results as well as with dielectric measurements the impact of the degree of H bonding on the magnitude of the electric polarization and thus dissipation is clarified.

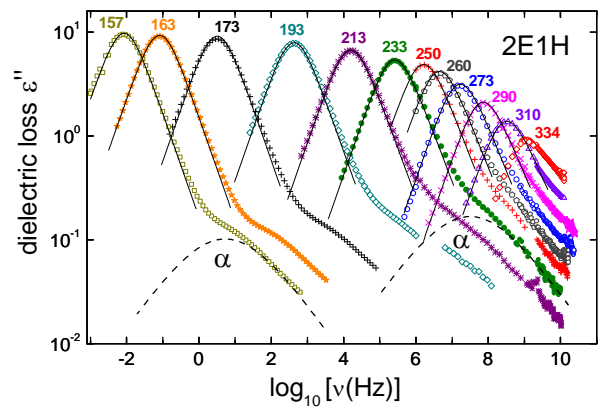

Fig. 1 (Color online) Dielectric loss $\varepsilon''(\nu)$ for 2-ethyl-1-hexanol in an extended temperature and frequency range. The numbers indicate temperatures in Kelvin. The solid lines highlight the Debye process and the dashed lines represent contributions of the α-process to the spectra measured at 157 K and 233 K.

To gain an impression of the latter, in Fig. 1 we present the absorption of 2E1H, in terms of the dielectric loss $\varepsilon''(\nu)$ which was recorded over more than 13 decades in frequency [12], thus extending the scope of previous measurements substantially. The data exhibit a prominent Debye-like profile, $\varepsilon''(\nu) = \Delta\varepsilon_D \, 2\pi\nu\tau_D / [1 + (2\pi\nu\tau_D)^2]$, represented in Fig. 1 as solid lines. Its amplitude or, more precisely, its dispersion strength $\Delta\varepsilon_D$ is much larger and its relaxation time $\tau_D$ much longer than the corresponding quantities, $\Delta\varepsilon_\alpha$ and $\tau_\alpha$, characterizing the α-process. The latter was identified previously by comparison with thermodynamic and viscoelastic data [6,7] and its contribution to the dielectric spectra is highlighted in Fig. 1 by dashed lines for two temperatures.

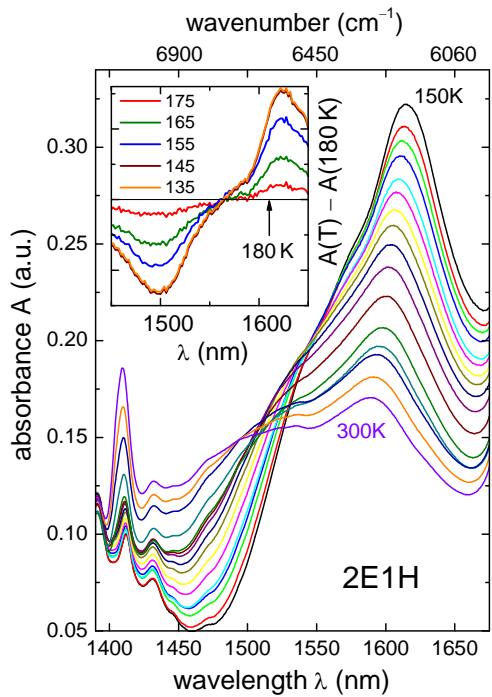

Fig. 2 (Color online) The overtones of the stretching modes corresponding to the H bonded (at high λ) and non-bonded (low λ) hydroxyl groups in 2E1H as measured from 300 K down to 150 K in steps of 10 K. The difference spectra in the inset are obtained by subtraction from the reference spectrum measured at 180 K. At 145 and 135 K, i.e., in the glassy state, the difference spectra are practically indistinguishable.

Focusing on the Debye relaxation, one recognizes that its peak amplitude depends only weakly on temperature for T ≤ 250 K, but decreases more rapidly for higher T. The observed strong decrease of the relaxation strength can be suspected to arise from a successive breaking of hydrogen bonds at high temperatures, corroborating the notion that H bonds play a prominent role in the generation of the Debye process. An immediate handle on the potentially underlying H bond equilibrium is provided most directly by virtue of temperature dependent NIR experiments [13] if one exploits that the frequency of the OH stretching vibrations [14] or their overtones [15] sensitively reflect changes in the molecular environment. The first overtones of OH stretching bands usually appear at wavelengths λ between 1400 and 1650 nm, with the free OH groups vibrating close to 1400 nm and strongly H bonded OH groups contributing close to 1600 nm [16].

NIR spectra of 2E1H are shown in Fig. 2 in the relevant wavelength range. At room temperature we find a well-defined absorbance maximum, $A_{1410}$, close to 1410 nm which corresponds to free, non-hydrogen bonded OH groups. The series of peaks between 1450 and 1650 nm signals a distribution of H bonded states that can be ascribed to OH stretch overtones of different aggregates [17]. The absorbance in Fig. 2 reveals an enormous T dependence in the entire λ range. With decreasing temperatures the fraction of free OH groups decreases significantly and concomitantly the distribution of H bonded OH groups becomes narrower leading to a well-defined and steadily growing maximum near 1620 nm. Close to the glass transition temperature $T_{g,cal}$ (≈ 146 K for 2E1H [6]) only a small fraction of free OH groups and a large fraction of strongly H bonded states is found. The shift of the main absorption to larger wave lengths on decreasing temperatures indicates a redistribution of the H bonded states, governed by an increasing strength of the predominant H bonds [16]. Focusing on T ≤ 180 K the differential NIR absorbance $\Delta A(\lambda,T) = A(\lambda,T) - A(\lambda,180K)$ shown in the inset of Fig. 2 looks as if an isosbestic point exists [18].

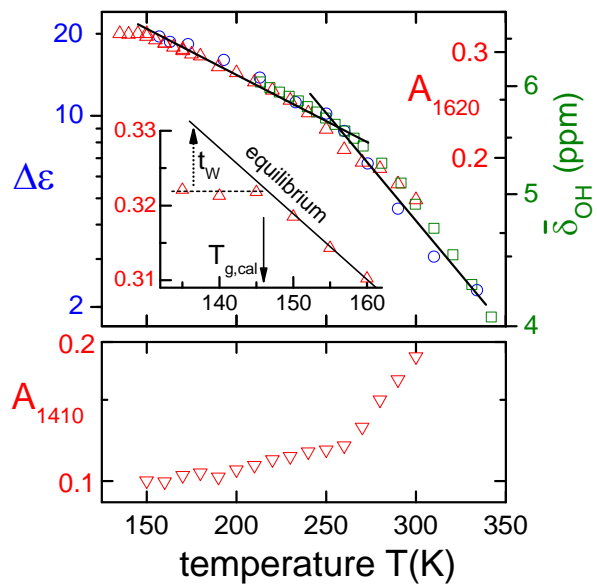

Fig. 3 (Color online) Changes in the temperature dependence are observed around 250 K in the dielectric strength $\Delta\varepsilon_D$ of the Debye process (circles) and in metrics of the equilibrium population of the H bonded hydroxyl groups, as obtained via the maxima of the absorbance peaks near 1410 and 1620 nm, (triangles), and via the chemical shift $\overline{\delta}_{OH}(T)$ of the hydroxyl protons, (squares). The inset depicts $A_{1620}(T)$ near $T_g$ demonstrating that for $T < T_g$ the absorbance is "frozen in" on the time scale of the corresponding experiments. The dashed arrow is intended to indicate that for sufficiently long waiting times $t_w$ the peak absorbance may return to its extrapolated equilibrium value (solid line).

To check whether the T dependence of the intact versus broken H bonds is correlated with the dielectric strength of the Debye relaxation, Fig. 3 (upper frame) compares $\Delta\varepsilon_D(T)$ with the dependence of the absorbance maximum close to 1620 nm, $A_{1620}$, The lower frame of Fig. 3 documents the temperature evolution of the free OH groups via the maximum of the absorbance close to 1420 nm (given as 1 – $A_{1420}$). All three quantities reveal a clear change of slope close to 250 K unambiguously demonstrating that the strength of the Debye process reflects indeed the fraction of intact H bonds.

Further evidence regarding the H bond equilibrium was obtained via NMR by tracing the isotropic chemical shift (CS) variation $\delta(T)$ of the alcohol's hydroxyl proton. It is known [19,20] that protons in bonded (b) and in non-bonded (n) OH groups can exhibit a significant CS difference $\Delta\delta_{OH} = \delta_n - \delta_b$. Obviously, the fraction $f_n$ of open H bonds (or closed ones, $f_b = 1 - f_n$) changes with temperature. Separate peaks at $\delta_n$ and $\delta_b$ with relative strengths of $f_n$ and $1 - f_n$, respectively, are, however, not resolved experimentally as long as the H bond switching occurs with rates $1/\tau_{ex}$ larger than $|\Delta\delta_{OH}|$ [19]. In this fast-exchange limit, which applies through most of the supercooled regime, only a motionally averaged NMR line shows up at a CS of $\overline{\delta}_{OH}(T) = f_n\,\delta_n + f_b\,\delta_b$. Tracing $\overline{\delta}(T)$ against a shift standard thus allows one to extract $f_n$, or its variations, experimentally. Advantageously, the $CH_3$ groups in alcohols with their virtually T independent CS can be utilized as an internal standard [19,20,21]. The shifts $\overline{\delta}_{OH}(T)$ thus obtained for 2E1H in a magnetic field of 7.05 T, also included in the upper frame of Fig. 3, independently confirm the discussed temperature dependent H bond equilibrium.

It is worthwhile to observe in Fig. 2 that the difference NIR spectra for 135 and 145 K are identical. They were recorded in thermal equilibrium subsequent to cooling the sample at an average rate of 1.6 K/min. It should be realized that if H bond equilibration requires the existence of molecular rearrangements it should effectively be frozen below $T_{g,cal}$. In fact, Fig. 3 reveals that the absorbance $A_{max}$ remains virtually unchanged for T < 150 K and the given thermal history. Since the glass transition is a kinetic phenomenon, aging the sample somewhat below $T_{g,cal}$ for sufficiently long waiting times, $t_w$, will nevertheless lead to a restoration of equilibrium. In order to determine the H bond exchange time $\tau_{ex}$ we performed temperature down-jump experiments during which the NIR absorbance $\Delta A(t_w)$ was monitored [22]. Subsequent to a temperature step of $\Delta T = 3$ K our setup permitted to stabilize the base temperature within ~ 60 s. The adjustment of the H bond equilibrium, which is coupled to the liquid dynamics, took much longer: For a base temperature of 142 K the inset of Fig. 4 presents $\Delta A_n(t_w)$ (normalized to decay from 1 to 0) which can be described by a time constant $\tau_{ex} = 850$ s. As the relaxation map, Fig. 4, shows, $\tau_{ex}$ is consistent with the structural relaxation time $\tau_\alpha$ but much shorter than the time constant of the dominating Debye process.

There are widespread notions as well as models assuming that the Debye process can be regarded as a direct consequence of H bond switching [3,4,10,11,23]. The dynamics of the latter is directly detected by the present time-dependent NIR experiments. Thus, within this scenario, $\tau_{ex}$ deduced from these experiments should agree with $\tau_D$. However, this obviously is not the case (Fig. 4), which provides clear evidence that the Debye process cannot directly correspond to the H bond dynamics. Instead, $\tau_{ex}$ matches $\tau_\alpha$. This finding can be rationalized by considering that the $\alpha$-dynamics is directly related to the viscous flow [24,25] which, as a prerequisite, requires the breaking of H bonds [26].

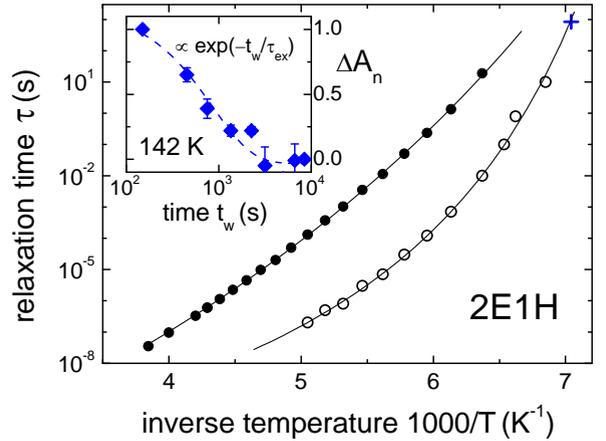

Fig. 4 (Color online) Temperature dependent dielectric relaxation times for the Debye process (closed circles) and the $\alpha$-process (open circles). The solid lines are fits using the Vogel-Fulcher equations $\tau_D/s = 4\times10^{-15}\exp[3020\,K/(T-73\,K)]$ and $\tau_\alpha/s = 4\times10^{-13}\exp[1170\,K/(T-109\,K)]$. The cross marks the time, $\tau_{ex} = 850$ s, needed for the equilibration of the measured NIR spectra. The inset shows the time variation of the normalized NIR absorbance at 142 K (diamonds) with the dashed line representing an exponential decay.

While Fig. 4 demonstrates that the Debye peak is not directly caused by the forming and breaking of H bonds, the results presented in Fig. 3 reveal a close connection of both phenomena. Such a connection is also implicit in a recently suggested transient chain model (TCM) which explains the exponentiality of the Debye process and furthermore features H bond reformation on a time scale $<< \tau_D$ [27,28]. The TCM is based on the NMR detection of an OH relaxation time $\tau_{OH} \geq \tau_\alpha$ for T >> $T_g$ while dielectric data indicate that $\tau_D$ (>> $\tau_{OH}$) approaches $\tau_\alpha$ upon cooling close to $T_g$ [29,30]. Within the TCM the approach of $\tau_{OH}$ and $\tau_\alpha$ is in harmony with the present NIR observation made near $T_g$ and moreover it explains why physical aging does not, as usual, take place on the slowest scale time scale in the system (here $\tau_D$) but on the scale set by $\tau_\alpha$ [31].

The change in the H bond equilibrium, obvious from the change in slope in Fig. 3, is not anticipated from the behavior of H bonded liquids devoid of a Debye process, such as the polyalcohol sorbitol [15]. But, this change is compatible with the TCM [27] because within that model the transient chains are destabilized as temperature (or pressure [32]) increases, leading to a reduction of their end-to-end distance. On the basis of dielectric experiments such a destabilization can not only be recognized from a decrease of $\Delta\varepsilon_D$ with increasing T but also from the approach of the time scales $\tau_\alpha$ and $\tau_D$ in the high-T regime [24,29].

To summarize, wide temperature range NIR and NMR spectroscopy in a monohydroxy alcohol was used to map out its H bond equilibrium population. We find that it parallels the dielectric strength that characterizes the prominent Debye-like dielectric absorption peak. Despite this coincidence regarding the static properties, the equilibration of the H bond population proceeds much faster than anticipated from the time scale associated with the Debye peak. This documents that its origin can not be regarded as a direct consequence of H bond switching as was previously believed, and is in accord with a recently advanced model.

We thank R. Gainaru and A. Sapsford for technical assistance and H. H. Limbach for interesting discussions. Support of this project by the Deutsche Forschungsgemeinschaft under Grant No. BO1301/8-1 and via Research Unit FOR1394 is gratefully acknowledged.

———————————